\documentclass[prl,twocolumn,showpacs,preprintnumbers,amsmath,amssymb]{revtex4}
\usepackage{graphicx}
\usepackage{latexsym}
\usepackage{color}

 \newcommand{\bB}{\mathbf{B}}

\newcommand{\bj}{\mathbf{j}}

 \newcommand{\bv}{\mathbf{v}}

\begin{document}
\title{Subcritical dynamo bifurcation in the Taylor Green flow}

\author{Y. Ponty$^1$, J.-P. Laval$^2$, B.
Dubrulle$^3$, F. Daviaud$^3$, J.-F. Pinton$^4$}
\affiliation{$^1$ Laboratoire Cassiop\'ee,  CNRS
UMR6202, Observatoire de la C\^ote d'Azur, BP
4229, Nice Cedex 04, France \\
$^2$ Laboratoire de M\'ecanique de Lille, CNRS
UMR8107,  Bd P. Langevin, 59655 Villeneuve d'Asq,
France \\
$^3$ Service de Physique de l'Etat Condens\'e, CNRS 
URA 2464, CEA Saclay, 91191 Gif-sur-Yvette, 
France\\
$^4$  Laboratoire de Physique, de l'\'Ecole Normale
Sup\'erieure de Lyon, CNRS UMR5672, 46 All\'ee d'Italie, 69007 Lyon, France}

\begin{abstract}
We report direct numerical simulations of dynamo generation
for flow generated using a Taylor-Green forcing.
We find that the bifurcation is subcritical,
and show its bifurcation diagram.
We connect the associated hysteretic behavior
with hydrodynamics changes induced by the action of the Lorentz force.
We show the geometry of the dynamo magnetic field and
discuss how the dynamo transition can be induced 
when an external field is applied to the flow.
\end{abstract}
\maketitle

Larmor~\cite{larmor} is generally credited for 
suggesting that the magnetic field
of the Sun (and, by extension, that of planets and other celestial bodies)
could be due to dynamo action -- {\it 
i.e.} self-generation from the motions
of an electrically conducting fluid. This principle has received much
theoretical support~\cite{theory} since then and has recently been validated
by experimental observations~\cite{Riga,Karlsruhe,vksP1,vksP2}.
Dynamo action results from an instability: when the flow
magnetic Reynolds number $R_M$ exceeds a critical value $R_M^c$,
the null magnetic field state looses its stability to a non-zero 
magnetic field state. Because of the low value the magnetic Prandtl number
of the considered fluids, this instability usually happens on a 
turbulent (noisy) basic state and the choice of an order parameter
can be ambiguous~\cite{bruit}. However, we can assume that the usual concepts of 
stability theory apply (cf. later) and study if the 
transition is supercritical or subcritical~\cite{subNL}.
In most models and in all experiments, this bifurcation is supercritical:
$R_M^{c}$ is a unique number, albeit flow 
dependant. For instance $R_M^c \sim 14$  and 
$R_M^c \sim 18$ for the constrained Karlsruhe and 
Riga experiments, while $R_M^c \sim 32$ for the fully
turbulent VKS dynamo~\cite{vksP1}.
On the other hand, the dynamo bifurcation may 
also be subcritical particularly because the action 
of a growing magnetic field is supposed to reduce hydrodynamic 
turbulence and maintain dynamo action for lower 
$R_M$ values. In fact, the transition can be globally subcritical
if the basic state experiences instability with respect to finite 
amplitude perturbations~\cite{DauchotManneville}.
A characteristic hysteretic behavior is then associated to the 
bifurcation, and the dynamo operates for a range of lower values  
$R_M^g < R_M < R_M^c$. Subcriticality has been 
discussed in MHD Alpha-Omega dynamical systems~\cite{subalpha1,subalpha2} 
and also for numerical simulations of convective dynamos in 
spherical geometries~\cite{morin}

In this Letter, we study the dynamo bifurcation  using full MHD 
simulations, generated in a 3D-periodical domain, 
by the Taylor Green forcing~\cite{brachet}.  At 
low Reynolds numbers, this flow has several 
metastable hydrodynamics 
states~\cite{dubrulle07}. At higher Reynolds 
numbers, it has a well defined mean flow 
structure with superimposed intense turbulent 
fluctuations. Recent studies of the linear 
problem have shown that, while the dynamo 
thresholds may runaway in flows generated by 
random forcing~\cite{runaway}, a dynamo is 
observed at all kinetic Reynolds 
numbers~\cite{ponty05,ponty07,laval06} in the 
Taylor-Green flow. We study the fully nonlinear 
regime and report here evidence of the 
subcriticality of the bifurcation.

Using standard direct numerical simulation (DNS) 
procedures we integrate pseudospectrally the MHD 
equations in a $2\pi$-periodic box:
\begin{eqnarray}
&&\frac{\partial {\bf v}}{\partial t}+ {\bf v} \cdot \nabla {\bf v} =
-\nabla {\cal P} + \bj \times \bB +\nu \nabla^2 \bv + {\bf F}, \label{E_MHDv}\\
&&\frac{\partial {\bf B}}{\partial t}+ {\bf v} \cdot \nabla {\bf B} =
\bB \cdot \nabla {\bf v}
+\eta \nabla^2 {\bf B} \ ,
\label{E_MHDb}
\end{eqnarray}
together with ${\bf \nabla} \cdot {\bf v} = 
\nabla \cdot {\bf B} =0$; a constant mass density 
{$\rho = 1$} is assumed. Here, $\bv$ stands for 
the velocity field, $\bB$ the magnetic field (in 
units of Alfv\'en velocity), $\bj=(\nabla \times 
\bB)/\mu_0$ the current density, $\nu$ the kinematic viscosity,
$\eta$ the magnetic diffusivity and ${\cal P}$ is 
the pressure.  The forcing term ${\bf F}$ is 
{given by} the TG vortex
\begin{equation}
{{\bf  F}_{\rm TG}(k_0)}= { 2f(t) } \,  \left[
\begin{array}{c}
\sin(k_0~x) \cos(k_0~y) \cos(k_0~z) \\
- \cos(k_0~x) \sin(k_0~y) \cos(k_0~z)\\ 0
\end{array} \right]  \ ,  \label{eq:Ftg} \end{equation}
implemented here at $k_0=1$. In the sequel, we 
consider two types of forcing: one in which 
$f(t)$ is set to a constant -- this is the 
constant force forcing ($f(t)=1.5$) considered 
in~\cite{ponty05}. In a second one $2f(t)$ is set 
by the condition that the $(1,1,1)$ Fourier 
components of the velocity remains equal to the 
Taylor-Green vortex -- this is the constant 
velocity
forcing considered in \cite{laval06}. For the 
linear instability problem, both forcing yield 
the same value of $R_M^c$~\cite{ponty05,laval06}. 
We now explore the non-linear regime, as well as 
the response to finite amplitude perturbations. 
Three control parameters drive the instability: 
the magnetic and kinetic Reynolds numbers and the 
amplitude of an external magnetic field $B_0$ when 
applied.
\begin{equation}
R_M=\frac{v_{\rm rms}^{0} \pi}{\eta} \; \; \;
R_V=\frac{v_{\rm rms}^{0} \pi}{\nu} \; \; \;
\Lambda = \frac{B_0}{v_{\rm rms}^{0}} \ .
\end{equation}
In the definition of the Reynolds numbers, the 
characteristic length scale is set to $\pi$, the 
size of a TG cell when $k_0=1$. The 
characteristic speed $v_{\rm rms}^{0}$ is 
computed from hydrodynamic runs in which the 
Navier-Stokes equation is not coupled to the 
induction equation,  $v_{\rm rms}^{0} = \langle 
\sqrt{2 E_V(t)} \rangle_t$. Here $E_V$ is net 
kinetic energy $E_V(t)$ and 
$\langle\cdot\rangle_t$ stands for averaging in 
time ($1/T~\int^T \cdot~dt$).  Likewise, in 
dynamo runs, the intensity of the magnetic field 
is estimated from the net magnetic energy 
$E_M(t)$, as $b = \langle \sqrt{2 E_M(t)} 
\rangle_t$.

\begin{figure}[h]
\centerline{\includegraphics[width=0.6\columnwidth]{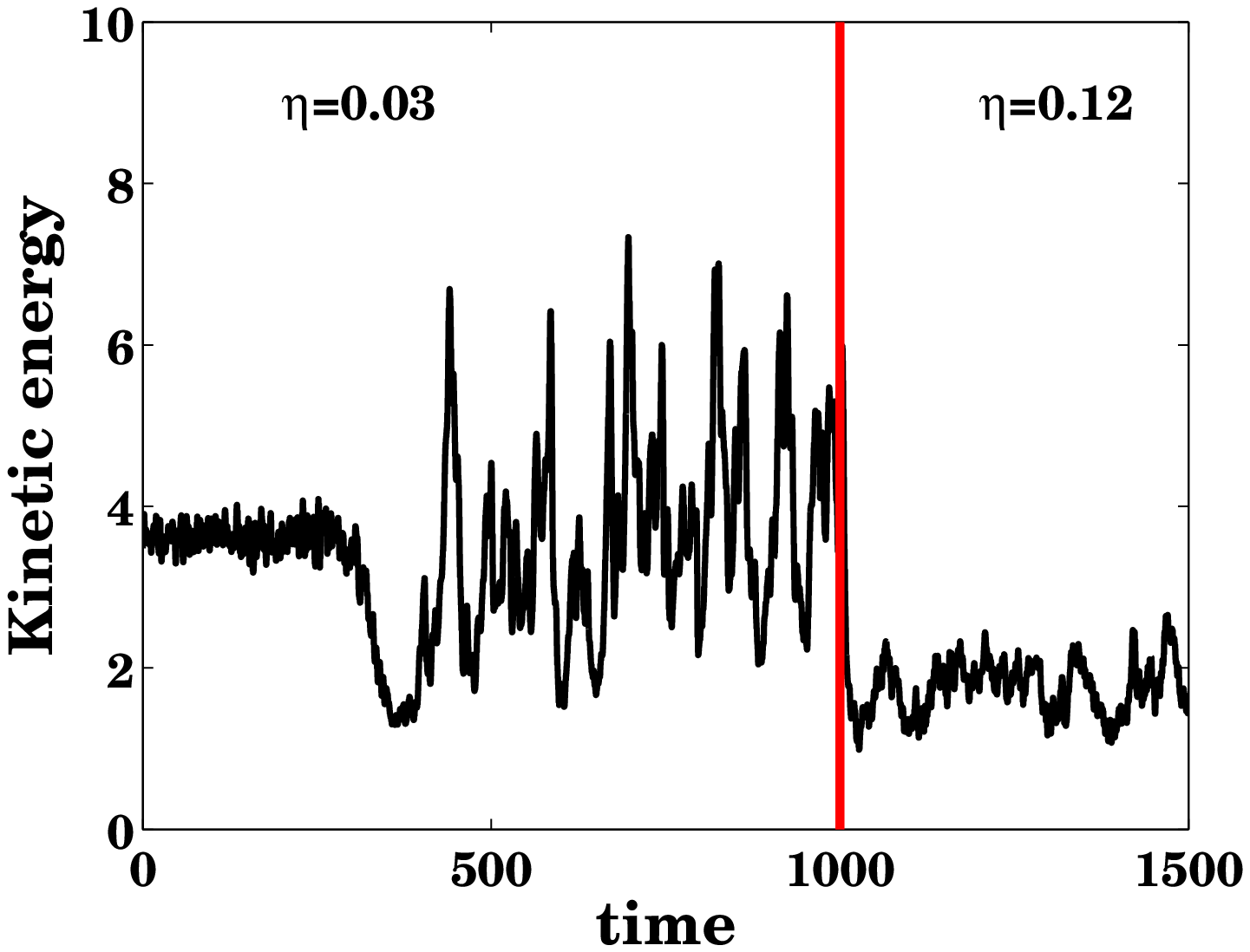}}
\centerline{\includegraphics[width=0.6\columnwidth]{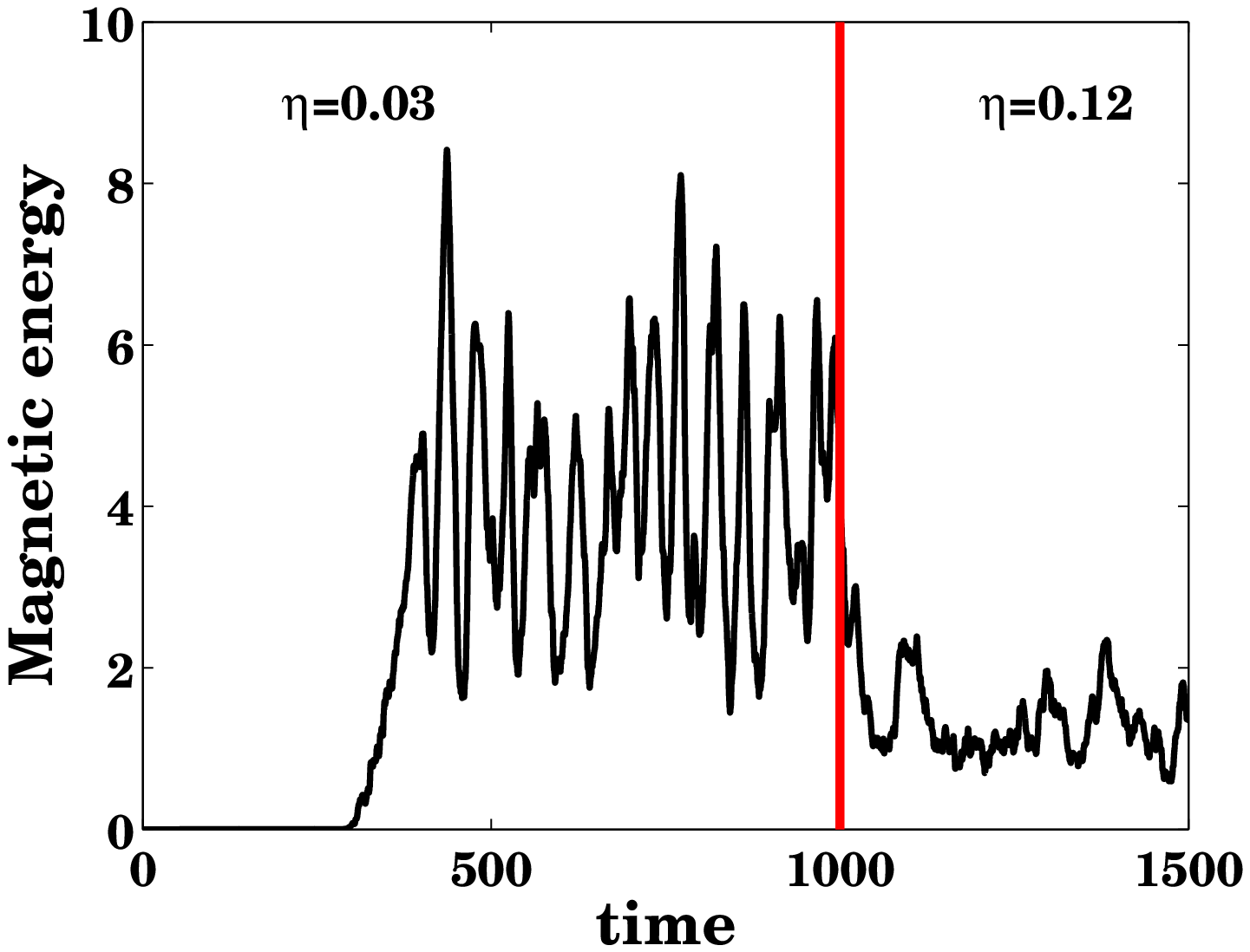}}
\caption{After a dynamo is self-generated from 
infinitesimal perturbations, the induction 
equation is quenched at $t= 1000$ by a four-fold increase of 
the magnetic diffusivity. It corresponds to a 
sudden change from $A$ to $A_9$ -- cf. Table I.}
\label{energies}
\end{figure}

Previous works~\cite{ponty07,laval06} have 
explored  the response of TG flows to 
infinitesimal magnetic perturbations, as a 
function of the  kinetic Reynolds number $R_V$. 
It was found that at any $R_V$, there exists a 
critical $R_M^c$ above which perturbations grow 
exponentially. This is illustrated in 
Fig.~\ref{energies} for a run at $R_V=563$ and 
$R_M= 281$ above the critical value 
$R_M^c=206$. The initial magnetic field 
perturbation -- with an energy level 
$E_M=10^{-17}$ -- first grows exponentially. At 
time $t\sim 300$, the magnetic field has reached 
sufficient
amplitude so  that it can react back onto the 
velocity field, saturate the instability and 
reach a statistically stationary state, with 
approximate equipartition $E_M\sim E_V$.
Note that times are given here in units of 
equation~(\ref{E_MHDv}), for which $1$ is very 
close to one eddy turnover time of the flow 
($T_{\rm NL}=\pi/v_{\rm rms}^{0} \sim 1.17$). 
This transition from infinitesimal perturbations 
builds the (solid) red curve in Fig.2.

We have then quenched the system: at $t=1000$, 
the magnetic diffusivity $\eta$ is suddenly 
increased by a factor of $4$, lowering $R_M$ 
below $R_M^c$. After a short transient, both 
$E_V$ and $E_M$ decrease and reach a second 
statistically stationary state, with a non zero 
magnetic energy -- a new dynamo state, for which 
equipartition is reached again 
(Fig.~\ref{energies}). This behavior is an 
evidence for global subcriticality~\cite{DauchotManneville}. The different 
levels of fluctuations in the two regimes suggest 
the possibility of different dynamo states, 
depending on the magnetic field or on history of 
the system.

\begin{table}[htb!]
\centerline{
\begin{tabular}{|c|c|c|c|c|c|c|c|c|c|} \hline
Point&$\eta$& $R_M$   &    $b$   & $L_B$   & $v_{\rm rms}$ &$L_U$  \\ \hline
$A$  & $0.03$ & $281$ &  $2.8$  & $5.2$  & $2.7$   & $3.0$   \\ \hline
$A2$ & $0.035$& $241$ &  $2.8$  & $5.3$ & $2.5$   & $3.0$   \\ \hline
$A3$ & $0.04$ & $211$ &  $2.8$ & $5.4$ & $2.5$   & $2.9$  \\ \hline
$A4$ & $0.05$ & $169$ &  $2.7$ & $5.5$ & $2.3$   & $2.9$  \\ \hline
$A5$ & $0.07$ & $121$ &  $2.6$ & $5.7$ & $2.0$   & $2.9$  \\ \hline
$A6$ & $0.08$ & $106$ &  $2.5$ & $5.7$ & $1.8$   & $2.9$  \\ \hline
$A7$ & $0.09$ & $94$  &  $2.4$ & $5.7$ & $1.7$   & $2.9$  \\ \hline
$A8$ & $0.1 $ & $84$  &  $2.0$ & $5.5$ & $1.7$   & $2.9$  \\ \hline
$A9$ & $0.12$ & $70$  &  $1.6$ & $5.1$ & $1.9$   & $3.0$  \\ \hline
$A10$ & $0.15$ & $56$  &  $0.0 $  & $0.0$   & $2.7$   & $2.6$  \\ \hline
\end{tabular}
}
\caption{For each regime: root mean square 
amplitude of the magnetic/{\it resp.}velocity 
fields $b = \langle \sqrt{2 Em(t)} \rangle$ \; , 
\;   $v_{\rm rms}=\langle \sqrt{2 Ev(t)} 
\rangle$, integral scale of the magnetic/{\it 
resp.}velocity fields $L_B = \langle \sum 
E_B(k,t))/k \rangle$ \; , \;  $L_U = \langle \sum 
E_V(k,t))/k \rangle$ --  $E(k,t)$ is the 
uni-dimensional energy spectra.}
\label{table:subdynamo}
\end{table}

\begin{figure}[h]
\centerline{\includegraphics[width=7cm]{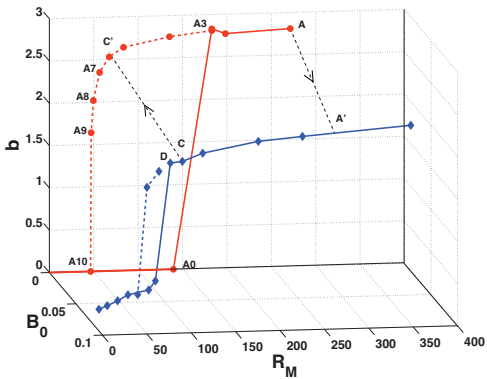}}
\caption{Bifurcation curves and hysteresis cycles 
when an external magnetic field is applied (full 
diamond symbols) or without one (full circle 
symbols). In this case, the subcritical quenched 
states (see text) form the red line. Jumps 
between the two branches link $A$ to $A'$ and $C$ 
to $C'$. }
\label{hysteresis}
\end{figure}

As subcritical bifurcations are also associated 
with hysteresis cycles, we have repeated the 
quenching  procedure starting from the same 
dynamo state $A$ (obtained at $t=1000$ at $R_V = 
563$ in Fig.~\ref{energies}) for increasing 
values of $\eta$, i.e. for {\it decreasing} $R_M$ 
values. The (time-averaged) magnetic and kinetic 
energy obtained after rearrangements are then 
recorded, and results summarized in 
Fig.~\ref{hysteresis} by the curve in the $B_0 = 
0$ plane.  Starting from  point A, one can 
sustain the dynamo after quenching through points 
$A2$ to $A9$, until a value $R_M^{g}$ 
substantially lower than $R_M^c$ (at $A9$, 
$R_M=70$ compared to $R_M=211$ in $A3$).

\begin{figure}[h]
\centerline{\includegraphics[width=4.6cm]{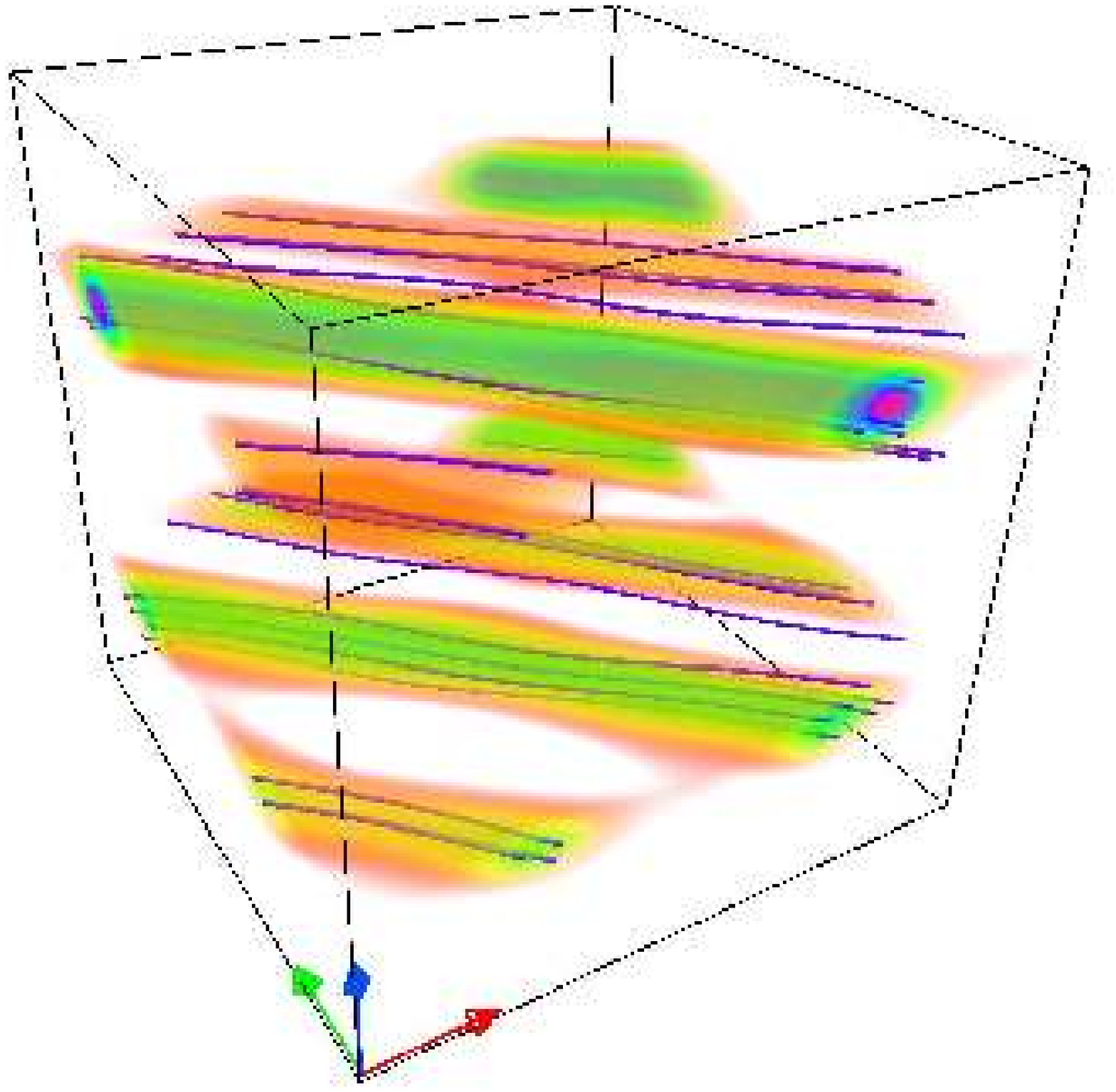}\includegraphics[width=4.1cm]{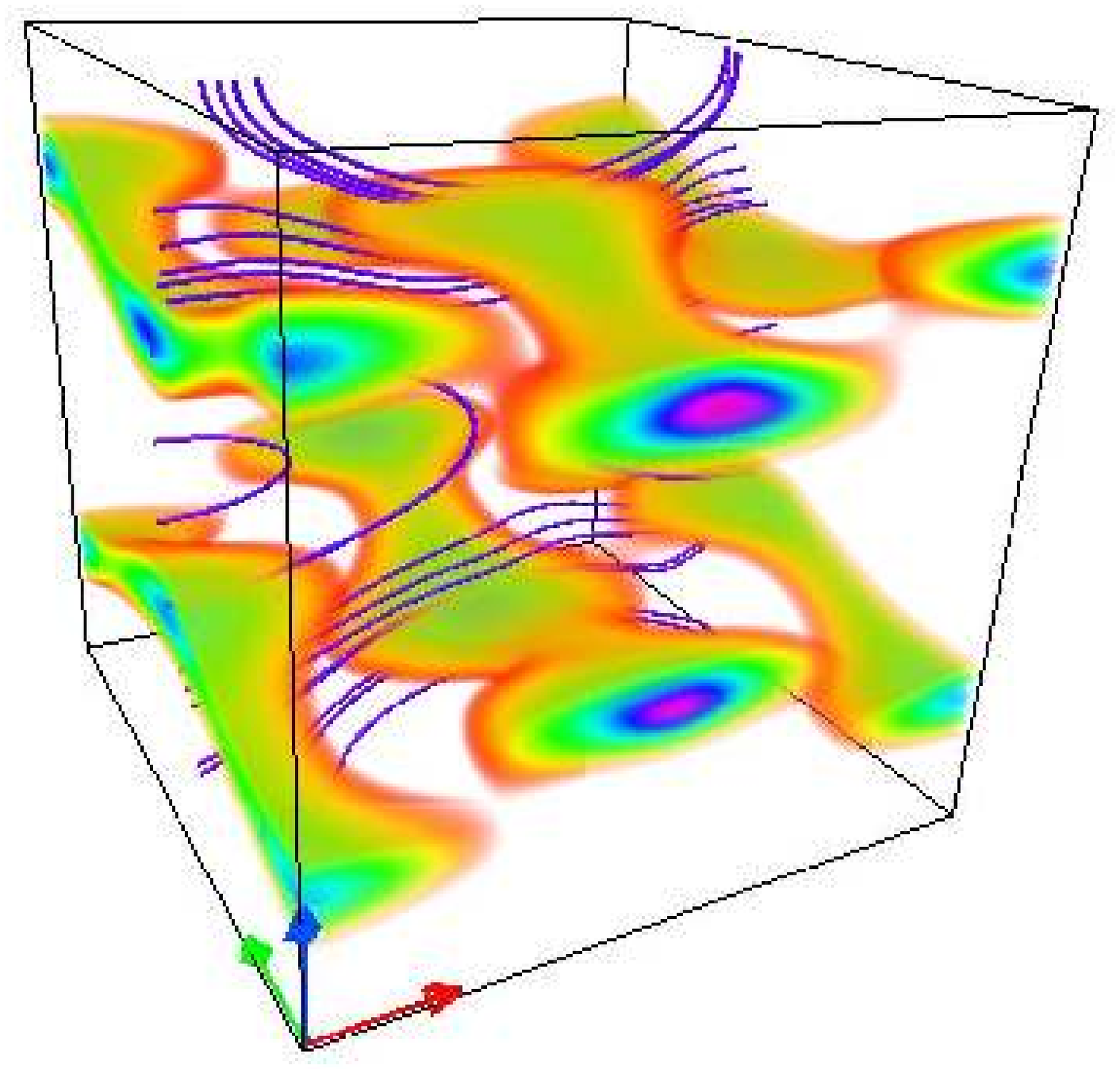}}
\caption {Volume rendering (75\% of ${\rm 
max}(b)$) of the magnetic energy and magnetic 
field lines~\cite{vapor}, for the normalized 
magnetic field $\langle B({\bf x}, t)/B(t) 
\rangle$ averaged in time during the run; 
(left) point $A7$ and (right)  point $A8$.}
\label{champmag}
\end{figure}

We have investigated further the system behaviour along the cycle by monitoring
the spatial structure of the magnetic and kinetic 
energies, so as to detect possible
changes in the flow structure.  In a first 
regime, until point $A7$, the kinetic energy (and 
hence $v_{\rm rms}$, i.e. the turbulence 
intensity -- see Table~I) decreases and so does 
the magnetic energy -- equipartition being essentially preserved.
Past $A8$, changes occur: $E_V$ starts to 
increase abruptly, while $E_M$ continues to 
decrease, resulting in a decreasing ratio 
$E_M/E_V$ -- see also Fig.4. Other global 
quantities are also changing along this branch 
(see Table I).  It corresponds to a modification 
in the spatial structure of the magnetic energy. 
As can bee seen in Fig.~\ref{champmag}, the 
dynamo modes in $A7$ and $A8$ are different. At 
$A7$, the dynamo has a structure with magnetic 
energy `tubes' in which the field line are 
concentrated along diagonal direction (aligned 
with the energy structures). In $A8$, the dynamo 
has a magnetic energy with a wavy shape and the 
field line are no longer parallel to the energy 
structures. In fact, the geometry of the $A8$ and $A9$ 
dynamo modes is reminiscent of the low kinematic 
mode of the TG dynamo\cite{ponty07}.

\begin{figure}[h]
\centerline{\includegraphics[width=8cm]{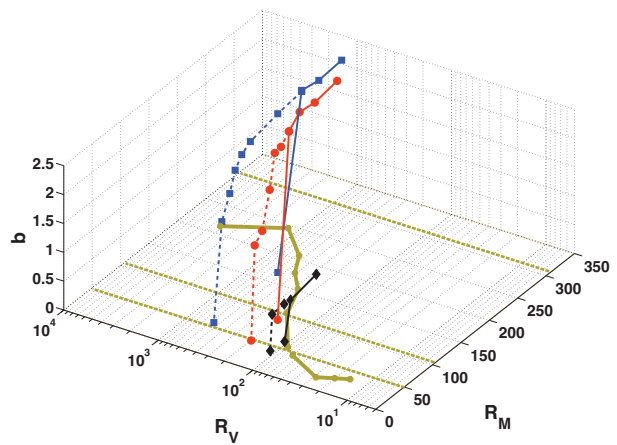}}
\caption{Hysteresis cycle for different Reynolds numbers and forcings --  constant force (red, blue) and constant velocity (black). The thick solid line in the $b=0$ plane is the linear instability $R_M^c$ {\it vs} $R_V$ from dynamical runs; the kinematic dynamo windows~\cite{ponty07}, $R_M \in [50, 110]$ and $R_M > 320$, are delimited by the thick dotted lines.}
\label{influrv}
\end{figure}

As turbulence influences the dynamo, we have 
repeated the above sequence of quenching at 
varying kinetic Reynolds numbers $R_V$. The 
result is shown in Fig.~\ref{influrv}.  We first 
observe that the hysteretic behavior persists as 
$R_V$ is lowered. In addition, the hysteresis 
cycle width,  $R_M^c - R_M^g$,  decreases with 
$R_V$.  It is interesting to compare their 
locations with respect to the dynamos windows 
evidenced in~\cite{ponty07,laval06} for the 
Taylor-Green forcing.  As shown in 
Fig.~\ref{influrv},  $R_M^g$ values are  almost 
independent of $R_V$ and lie close to the 
beginning of the first kinematic dynamo mode.  Of 
course, the onset $R_M^c$ switches from the 
kinematic low branch to the kinematic high branch 
as $R_V$ increases (and turbulence 
develops)~\cite{laval06,ponty07}. The width of 
the dynamo cycle is thus linked to the evolution 
of the $R_M^c(R_V)$ curve.

The above results were obtained with a constant 
force scheme. We have also repeated the quenching 
procedure using the constant velocity forcing. As 
can be seen in Fig.~\ref{influrv} (black curve / 
diamonds symbols), the hysteretic behaviour 
remains, but the transition towards the 
non-dynamo state is more abrupt. Another 
difference concerns the response to quenching; 
with a constant velocity forcing we observed a 
lower magnetic saturation level $b$. Those 
differences could be explained by a change in 
hydrodynamics properties such as the fluctuation 
level, at the same Reynolds number. In addition, 
when the velocity is kept constant there may be 
less possibility for the Lorentz force to change 
the flow.

Finally, we have checked the influence of finite 
amplitude external perturbations
on the hysteresis cycle by applying an external 
magnetic field of amplitude $B_0 = 0.07$ in 
the vertical direction. The result at $R_V=563$ 
is shown by the blue line in 
Fig.~\ref{hysteresis}. When comparing to the $B_0 = 0$
case (red curve), two effects 
are readily observed : (i) the hysteresis cycle 
is shortened and this is essentially due to a 
decrease in the onset $R_M^c$ from infinitesimal 
perturbations;  (ii) the amplitude of the 
magnetic energy in the dynamo is decreased, as 
lower $b$ values are obtained. These observations 
are indications that the external magnetic field 
has mediated a transition towards another 
equilibrium state~\cite{dubrulle07}. The 
transition towards this second equilibrium state 
is quite robust: one can also obtain it by 
switching on the vertical magnetic field starting 
from a state with a well-developed dynamo  (jump 
from $A$ to $A'$ in Fig.~\ref{hysteresis}). 
Conversely, starting from a dynamo state with an 
applied magnetic field and switching it off, one 
returns to the zero-magnetic field hysteresis 
curve
(jump from $C$ to $C'$ in Fig.~\ref{hysteresis}).

\begin{figure}[h]
\centerline{\includegraphics[width=8cm]{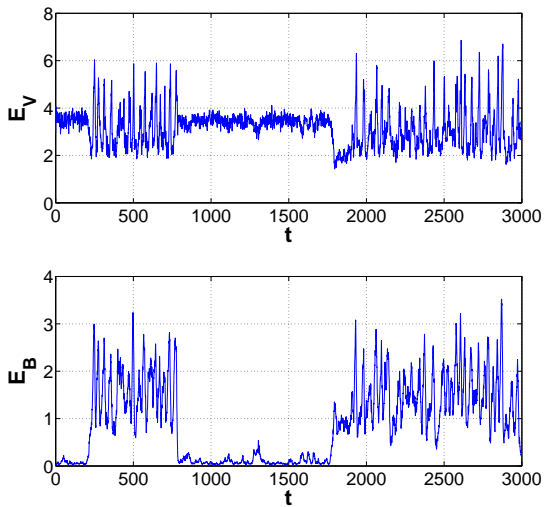}}
\caption{Evolution on time of the kinetic ($E_V$) 
and magnetic energy ($E_B$) when the flow is 
operated in the immediate vicinity of point $D$ 
-- see Fig.2.}
\label{onoff}
\end{figure}

A less deterministic behaviour is observed when the system is operated in the vicinity of point $D$ -- shown along the blue curve in Fig.2. At this point, the system is operated at a magnetic Reynolds number slightly smaller than the linear threshold ($93.8$ compared to about $100$) and one observes that the the systems spontaneously switches between dynamo and non-dynamo periods, as shown in  Fig.\ref{onoff}. This is reminiscent of the ``on-off'' bifurcation scenario sometimes proposed for the dynamo~\cite{FrickOnOff,LathropOnOff,LeprovOnOff,ForestOnOff} at high $R_V$. It has been observed in models~\cite{PlunianOnOff} and experimental~\cite{BVKpaper1} versions of the Bullard dynamo~\cite{Bullard}, and possibly in 
turbulent fluid dynamos~\cite{vksP2}. We note in Fig.\ref{onoff} that the kinetic energy has stronger fluctuations during the dynamo periods.


To summarize, we have evidenced in the TG flow 
several features characteristic of subcriticality of 
the dynamo instability. At variance with usual 
dynamical system,
this behaviour is obtained in a fully turbulent 
system, where fluctuations are of the same order 
of magnitude as the mean flow. We may remark that 
in this case, the traditional concept of 
amplitude equation may be ill-defined and one may 
have to generalize the notion of  `subcritical 
transition'  for turbulent flows. Another feature 
is the sensitivity to perturbations of the order 
parameter through the application of an external 
magnetic field. The perturbation mainly acts 
through macroscopic
changes in the system configuration (perturbation of the velocity field),
allowing lower thresholds for dynamo instability. 
These findings open new perspective for experimental dynamos. 
For the TG flow, we observe a decrease of  the 
dynamo threshold by as much as $57$ percent, 
with an external applied field of $B_0= 0.07$. 
We have also found that changes in the geometry of 
the dynamo states in the subcritical branch are 
consistent with the coexistence of several 
metastable hydrodynamics states~\cite{dubrulle07}. 
Preliminary observations in the VKS 
experiment also point to the existence of 
subcritical dynamos in the presence of global 
rotation~\cite{VKSsubcrit}, a feature also noted in some 
numerical models of the geodynamo~\cite{morin}.


\noindent {\bf Acknowledgements} We acknowledge 
useful discussions with A. Pouquet and P. 
Mininni, R. Jover and team members of the VKS 
collaboration. Computer time was provided by 
IDRIS and the Mesocentre SIGAMM at Observatoire 
de la C\^ote d'Azur. This work is supported by 
the French GDR Dynamo. YP thanks A. Minuissi for 
computing design assistance.


\begin{thebibliography}{99}

\bibitem{larmor}
J. Larmor, {\it Rep. Brit. Assoc. Adv. Sci}, 159-160, (1919).

\bibitem{theory}
H. K. Moffatt, ``Magnetic field  generation in 
electrically conducting fluids'',  (Cambridge U. 
Press, 1978);


\bibitem{Riga}
A. Gailitis et al.,
{\it  Phys. Rev. Lett.} {\bf 86}, 3024 (2001)

\bibitem{Karlsruhe}
R. Stieglitz and U. M\"uller,   {\it Phys. Fluids} {\bf 13} 561 (2001)

\bibitem{vksP1}
R. Monchaux et AD., {\it Phys. Rev. Lett.}  {\bf 98} 044502, (2007)


\bibitem{vksP2}
M. Berhanu et al., {\it Europhys. Lett.} {\bf 77}, 59001 (2007)

\bibitem{bruit}
R. Berthet et al., Physica {\rm 174D}, 84 (2003)

\bibitem{subNL}
P. Manneville, ``Dissipative Structures and Weak Turbulence'', 
(Academic Press, Boston, 1990)   

\bibitem{DauchotManneville}
O. Dauchot, P. Manneville, {\it J. Phys. II} {\bf 7}(2), 371 (1997)

\bibitem{subalpha1}
K. A. Robbins, {\it Proc. Nati. Acad. Sci. USA} {\bf 73}(12), 4297-4301 (1976)

\bibitem{subalpha2}
S. Fedotov, I. Bashkirtseva and L. Ryashko,
{\it Phys. Rev. E} {\bf 73}, 066307 2006.

\bibitem{morin}
V. Morin, Ph. D. Thesis, University Paris VI 
(1999) ; U.R. Christensen, P. Olson and G.A. 
Glatzmaier, {\it Geophys. J. Int.}, {\bf 138}, 
393 (1999).

\bibitem{brachet}
M. Brachet, {\it Fluid Dyn. Res.} {\bf 8}, 1 (1991);
C. Nore et al., {\it Phys. Plasmas} {\bf 4}, 1 (1997)

\bibitem{dubrulle07}
B. Dubrulle et al, {\it to appear in NJP} (2007).

\bibitem{runaway}
A.A. Schekochihin et al., {\it New J. Physics} {\bf 4}, 84 (2002);
A.A. Schekochihin et al., {\it Phys. Rev. Lett.}  {\bf 92}, 054502 (2004);
A. B. Iskakov et al., {\tt arXiv/astro-ph/0702291}

\bibitem{ponty05}
Y. Ponty et al., {\it Phys. Rev. Lett.} {\bf 94}, 164512 (2005)

\bibitem{laval06}
J.-P. Laval et al., {\it Phys. Rev. Let.} {\bf  96} 204503 (2006)

\bibitem{ponty07}
Y. Ponty et al.,  {\it  New J. Phys.} (2007)


\bibitem{vapor}
Imagery using VAPOR code ({\tt www.vapor.ucar.edu})

\bibitem{FrickOnOff}
S. Lozhkin, D. Sokoloff, P. Frick,  {\it Astronomy Reports} {\bf 43}(11), 753 (1999)

\bibitem{LathropOnOff}
D. Sweet et al.  {\it Phys. Rev. E} {\bf 63}, 066211 (2001)

\bibitem{LeprovOnOff}
N. Leprovost and B. Dubrulle, {\it Eur. Phys. J. B}, {\bf 44} 395 (2005).

\bibitem{ForestOnOff}
M. D. Nornberg et al.   {\it Phys. Rev. Lett.} {\bf  97}, 044503 (2006)

\bibitem{PlunianOnOff}
N. Leprovost, B. Dubrulle, F. Plunian,  {\it 
Magnetohydrodynamics} {\bf 42}, 131 (2006)

\bibitem{BVKpaper1}
M. Bourgoin et al.,  {\it New J.  Phys.} {\bf 8}, 329, (2006)

\bibitem{Bullard}
E. C. Bullard,  {\it Proc. Camb. Phil. Soc} {\bf 51}, 744 (1955)


\bibitem{VKSsubcrit}
VKS team, private communication.



\end{thebibliography}
\end{document}